\documentclass[sigconf]{acmart}
\usepackage{colortbl}
\usepackage{multirow}
\usepackage{caption}
\usepackage{subcaption}
\usepackage{xcolor}
\usepackage{comment}
\definecolor{myblue}{rgb}{135, 206, 235}
\definecolor{LightCyan}{RGB}{240,248,255}
\AtBeginDocument{%
  \providecommand\BibTeX{{%
    \normalfont B\kern-0.5em{\scshape i\kern-0.25em b}\kern-0.8em\TeX}}}

\copyrightyear{2024}
\acmYear{2024}
\setcopyright{rightsretained}
\acmConference[CHI EA '24]{Extended Abstracts of the CHI Conference on Human Factors in Computing Systems}{May 11--16, 2024}{Honolulu, HI, USA}
\acmBooktitle{Extended Abstracts of the CHI Conference on Human Factors in Computing Systems (CHI EA '24), May 11--16, 2024, Honolulu, HI, USA}
\acmDOI{10.1145/3613905.3636283}
\acmISBN{979-8-4007-0331-7/24/05}

\begin{document}

\title{With or Without Permission: Site-Specific Augmented Reality for Social Justice}


\author{Rafael M.L. Silva}
\orcid{0000-0001-8376-0416}
\email{rafaelsi@uw.edu}
\affiliation{%
  \institution{University of Washington}
  \country{USA}
}

\author{Ana María Cárdenas Gasca}
\orcid{0000-0003-4842-4729}
\email{acardenasgasca@ucsb.edu}
\affiliation{%
  \institution{UC Santa Barbara}
  \country{USA}
}

\author{Joshua A. Fisher}
\orcid{0000-0001-5628-5138}
\email{joshua.fisher@bsu.edu}
\affiliation{%
  \institution{Ball State University}
  \country{USA}
}

\author{Erica Principe Cruz}
\orcid{0000-0003-0992-0527}
\email{ecruz@cs.cmu.edu}
\affiliation{%
  \institution{Carnegie Mellon University}
  \country{USA}
}

\author{Cinthya Jauregui}
\email{cjauregui@scu.edu}
\orcid{0009-0008-2860-3264}
\affiliation{%
  \institution{Santa Clara University}
  \country{USA}
}

\author{Amy Lueck}
\orcid{0000-0002-0432-2280}
\email{alueck@scu.edu}
\affiliation{%
  \institution{Santa Clara University}
  \country{USA}
}

\author{Fannie Liu}
\orcid{0000-0002-5656-3406}
\email{fannie.liu@jpmchase.com}
\affiliation{%
  \institution{JPMorgan Chase \& Co.}
  \country{USA}
}

\author{Andrés Monroy-Hernández}
\orcid{0000-0003-4889-9484}
\email{andresmh@cs.princeton.edu}
\affiliation{%
  \institution{Princeton University}
  \country{USA}
}

\author{Kai Lukoff}
\email{klukoff@scu.edu}
\orcid{0000-0001-5069-6817}
\affiliation{%
  \institution{Santa Clara University}
  \streetaddress{500 El Camino Real}
  \country{USA}
  \postcode{95053}
}


\renewcommand{\shortauthors}{Silva, et al.}

\begin{abstract}
Movements for social change are often tied to a particular locale. This makes Augmented Reality (AR), which changes how people perceive their surroundings, a promising technology for social justice. Site-specific AR empowers activists to re-tell the story of a place, \textit{with or without} permission of its owner. It has been used, for example, to reveal hidden histories, re-imagine problematic monuments, and celebrate minority cultures. However, challenges remain concerning technological ownership and accessibility, scalability, sustainability, and navigating collaborations with marginalized communities and across disciplinary boundaries. This half-day workshop at CHI 2024 seeks to bring together an interdisciplinary group of activists, computer scientists, designers, media scholars, and more to identify opportunities and challenges across domains. To anchor the discussion, participants will each share one example of an artifact used in speculating, designing, and/or delivering site-specific AR experiences. This collection of artifacts will inaugurate an interactive database that can inspire a new wave of activists to leverage AR for social justice. 
\end{abstract}




\maketitle

\section{Introduction}
\begin{figure}[htbp]
  \includegraphics[width=0.5\textwidth]{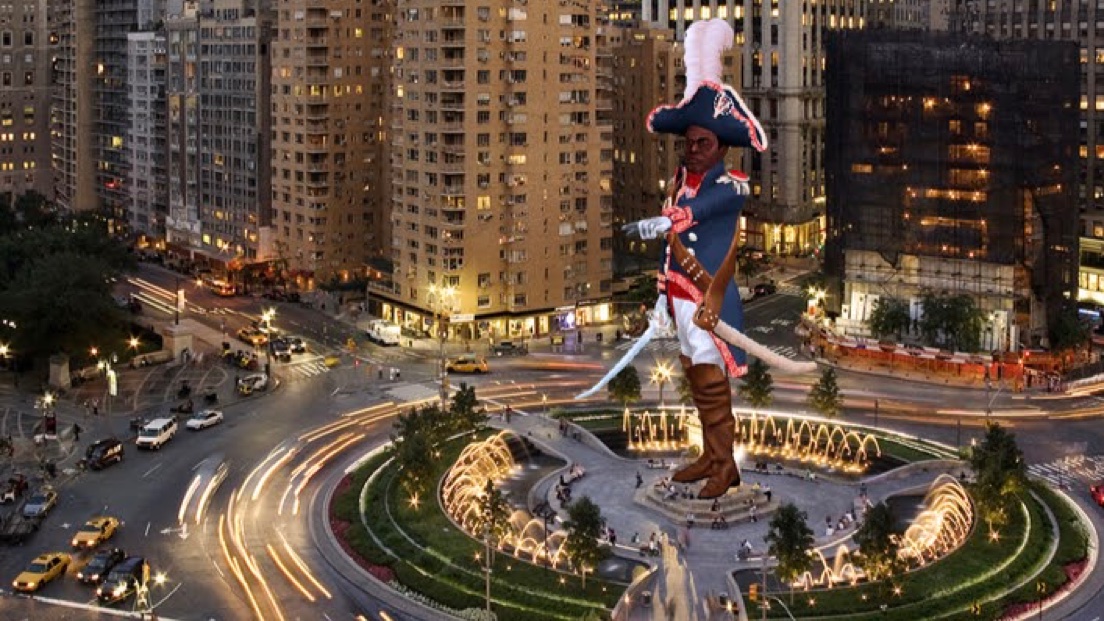}
  \caption{\textbf{An Example of a `Final Artifact.'} The Kinfolk App invites users to place digital monuments of people of color at specific locations. The project emerged from an artist and activist campaign to advocate for the removal of the statue of Christopher Columbus at Columbus Circle in New York City \cite{McWhirter2021-yj}. Here, Toussaint L'Ouverture, the leader of a Haitian slave rebellion and revolutionary movement, is shown superimposed over that statue of Columbus, leveraging augmented reality to envision a more just future.}
  \Description{Toussaint L'Ouverture is superimposed over Columbus Circle, a bustling traffic circle in New York City}
  \label{fig:kinfolk}
\end{figure}

Activism is often tightly connected with the places, spaces, and physical environments related to a cause \cite{UnderstandingARactivism}. With the rise of immersive technologies on everyday devices like smartphones, activists have created new ways of digitally augmenting physical locations to support social causes, leveraging the affordances of Augmented Reality (AR). Accordingly, activists have created AR experiences that re-evaluate the histories manifested in the social, cultural, and built environments \cite{MonumentLab,BLAM,Abebe2022-nq}. These include the Monument App, which invites users to place digital monuments of people of color at specific sites (Fig. \ref{fig:kinfolk}). 
In another example, SkytypingAR \cite{SkytypingAR} enabled activists to coordinate the display of protest messages condemning immigration policies above detention and correction centers. While such digital activism holds value in of itself, there is even precedent for it to serve as a `prototype' for changes that are later realized in the physical world \cite{Schroeder2023-qn}. 

However, open questions remain about when and how activists should leverage AR for social justice. For instance, when should activists work with existing powers, such as AR platforms that participate in surveillance capitalism, and when should they circumvent them? What methods can interdisciplinary collaborators use to co-design site-specific AR experiences? And how can activists reach wider audiences and maintain their creations?

\begin{figure}[htbp]
    \centering
    \includegraphics[width=0.6\textwidth]{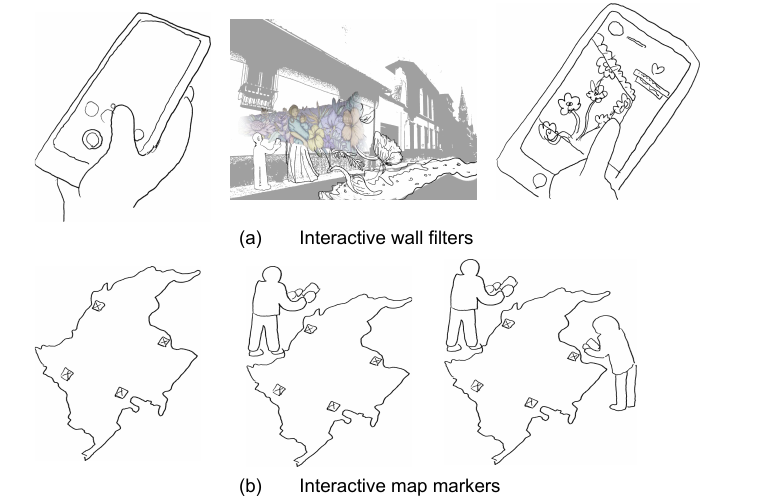}
    \caption{\textbf{An example of a ‘Speculative Artifact.'} The Memory Layers application, a depiction of sensitive narratives, was developed in collaboration with the Museo de la Memoria, a human rights museum in Colombia \cite{cardenas2022_2}. The research team and museum employees co-created sketches of three different modes of user interaction, focusing on adapting existing exhibitions and collections to an AR format. This exploration delved into a design space for AR applications, considering users, technology, and objectives.}
    \label{fig:speculativeartifact}
    \Description{Three rows of sketches, each displaying a different AR interaction. First row: interacting with a wall on the phone to show a mural in AR. Second row: multiple people scanning markers on a large map on the floor. Third row: person waving their arms to select between stories that are displayed on the wall.}
\end{figure}

\begin{figure}
 \begin{subfigure}[t]{0.49\textwidth}
                \includegraphics[width=\textwidth]{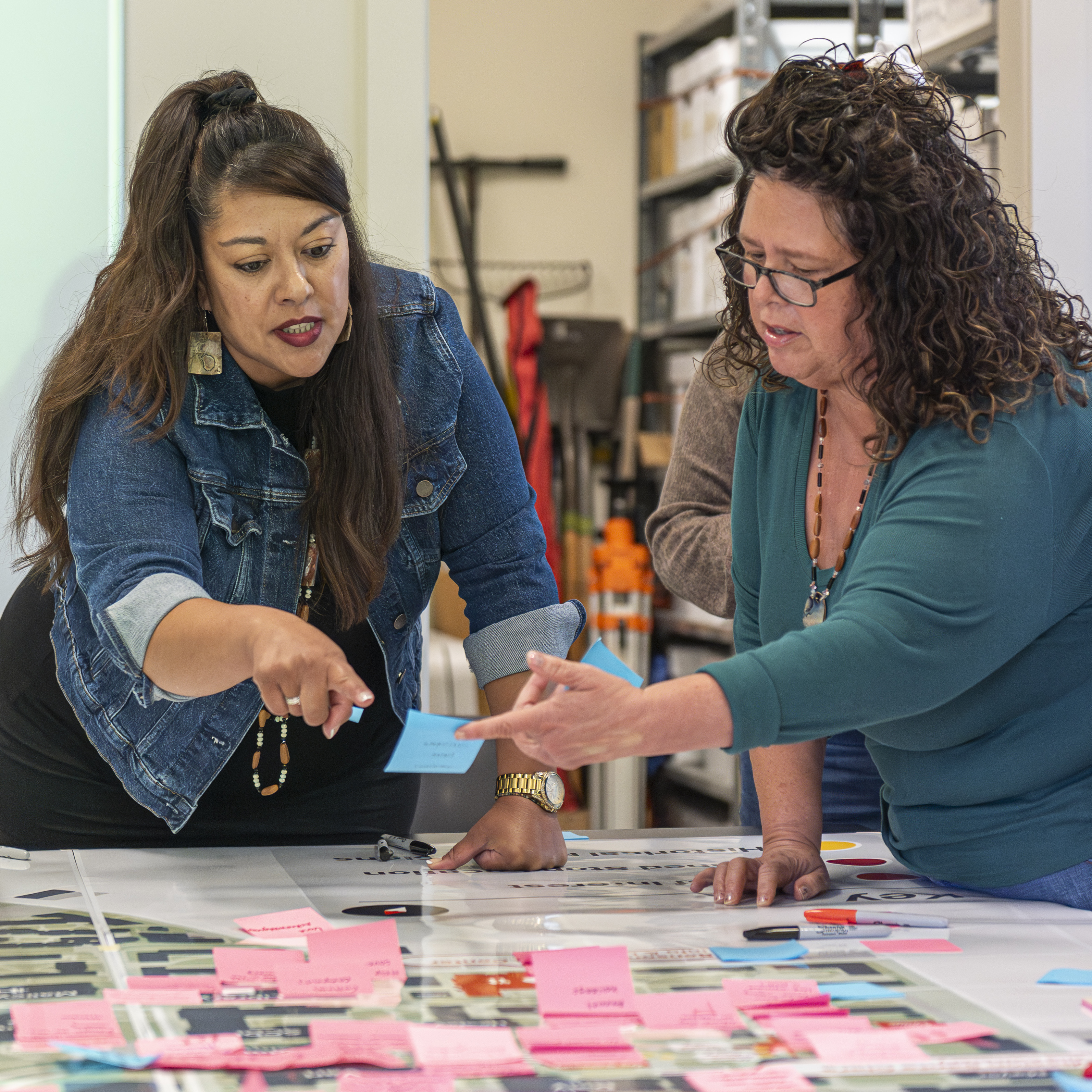}
                \caption{To create the Thámien Ohlone AR Tour at Santa Clara University, the team tried a new co-design exercise called “landmark-based affinity diagramming” \cite{Lukoff2023-do}. Here, members of the Muwekma Ohlone Native American Tribe discuss where to place a media asset on the map. }
                \Description{Two members of the Muwekma Ohlone lean over a table with a giant map, pointing and discussing where to place a Post-it note as part of this co-design exercise.}
        \end{subfigure}%
        \hfill
  \begin{subfigure}[t]{0.49\textwidth}
                \includegraphics[width=\textwidth]{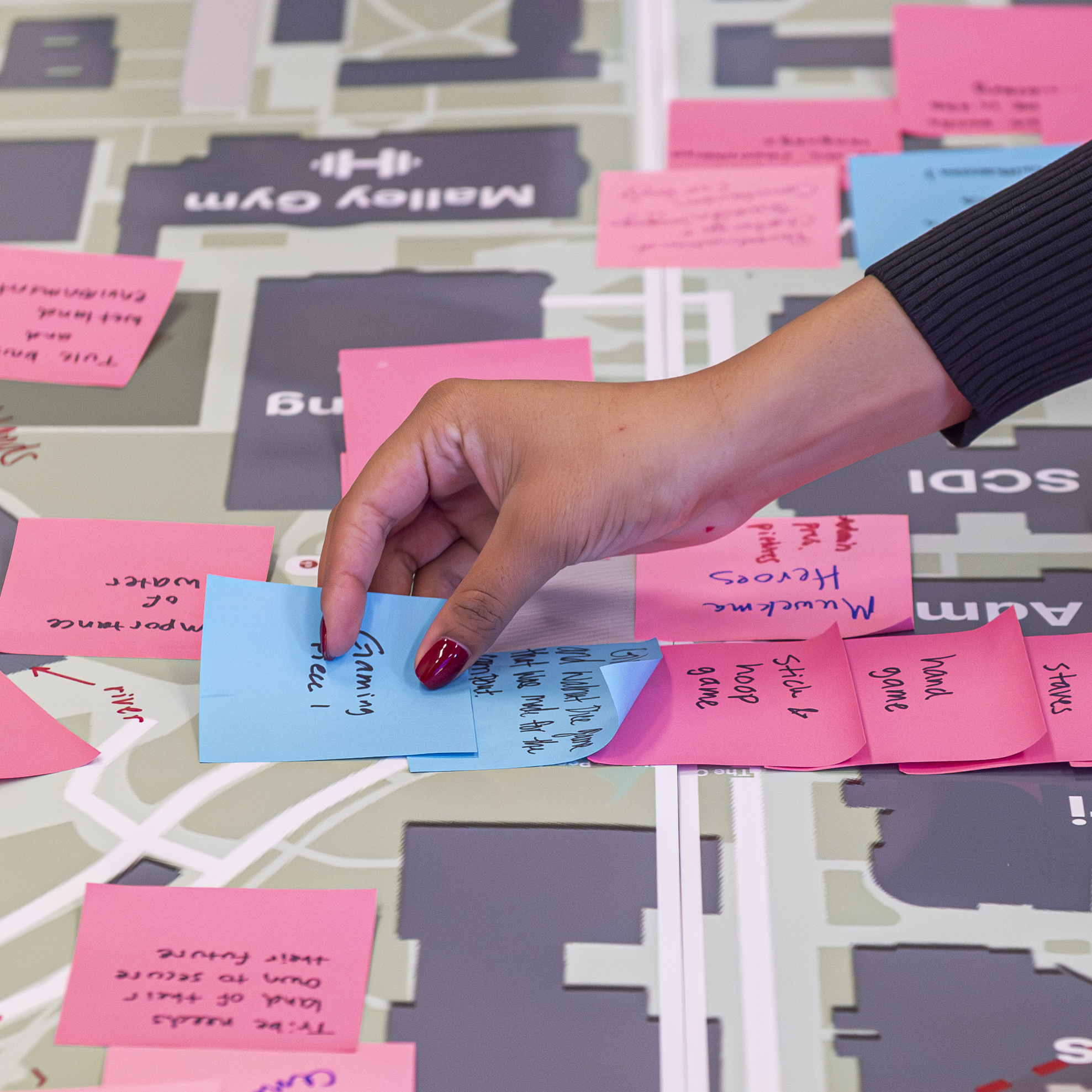}
                \caption{The team printed a 10 x 6 foot map of the historical campus, the site of the AR tour. Pink post-its are stories/messages the tribe wanted to share, e.g. “Hopes for a future Ohlone garden at the University.” Blue post-it notes are media assets, e.g. 3D models of tribal artifacts, recordings of songs, that were then displayed at that location in AR.}
                \Description{A hand places a Post-it note with hand-written text on a giant map of the Santa Clara University Campus. Many other Post-it notes have already been placed all over the map.}
        \end{subfigure}%
\caption{\textbf{An Example of a ‘Process Artifact’}}
\label{fig:processartifact}
\end{figure}

This workshop seeks to address these questions by centering  artifacts used in speculating, designing, and delivering AR experiences for social justice. Since it is difficult to explain AR experiences in words alone ~\cite{Bolter2021-bj}, these artifacts will ground the workshop discussions in concrete examples. One intended outcome of the workshop is an interactive database that will showcase these artifacts, serving as a repository that inspires future activists. Artifacts might include:
\begin{itemize}
\item \textit{A Speculative Artifact} such as a sketch, diagram, or description of an imagined future AR experience or creator tool (e.g., Fig. \ref{fig:speculativeartifact}).
\item \textit{A Process Artifact} such as a site map, an early prototype, a memorandum of understanding with a community partner, or a snippet of code (e.g., Fig. \ref{fig:processartifact}).
\item \textit{A Final Artifact} such as a 3D model, a soundscape, or a short video of the completed experience (e.g., Fig \ref{fig:kinfolk}).
\end{itemize}
The resulting collection of artifacts will facilitate discussion of a wide range of visions, methods, and results in the creation of AR experiences for social justice.

Drawing from Design Justice practice and theory~\cite{costanza2020design}, we see community engagement as the key to identifying locations of interest, connecting to existing social movements, and developing appropriate resources. We aim to translate our academic efforts into tools and knowledge that enable community-led praxis of AR for social justice. As AR technology goes mainstream, it becomes crucial to craft a future where AR does not become another hegemonic force but a tool to embolden prosocial change.

\section{Workshop Aims}
This workshop has three main aims:
\begin{itemize}
\item \textit{Build community around social justice AR}. Connect creators and researchers working in this emerging domain.
\item \textit{Interdisciplinary discussion}. Bring together activists, computer scientists, designers, media scholars, and more to identify opportunities and challenges for social justice AR across domains.
\item \textit{Develop an interactive database of artifacts}. Given the difficulty of accessing site-specific AR experiences, a repository of exemplars can serve as a resource for future social justice AR projects. 
\end{itemize}

\section{Background and Motivation}

\subsection{Opportunities and Challenges}
Place-based action has always been an integral aspect of social justice work and, as a technology tightly couple with physical world, AR introduces interesting affordances for this type of initiatives. In recent years, the widespread adoption of smartphones with AR technology has spurred activists to digitally augment places to support their causes. Practitioners can now overlay digital content onto the environment to embody their message. For example, users of the BeHere/1942 mobile AR exhibition app \cite{fujihataBeHere1942New} can witness the physical space of a Los Angeles block overlaid with a scene from the Japanese-American internment. GPS-enabled applications, such as the Emmett Till Memory Project \cite{instuteofmuseumandlibraryservicesRememberingEmmettTill}, harness the power of location to reveal neglected stories. By preserving the sites and narratives of the Till lynching, this project offers users an immersive connection to specific locales, emphasizing their historical significance.


The creators of these apps are confronted with a selection of frameworks and tools that influence not only the user experience but also touch upon political and ethical matters. Activists must consider what it means to use technologies that could be antithetical to their causes or introduce privacy, accessibility, and maintenance concerns. This workshop aims to start an interactive database that shares and categorizes annotated examples in order to help activists navigate this landscape.

\subsection{Outcomes and Future Speculations}

As a process and action, the practice of social justice using AR seeks to address a complex social challenge or issue. While there are no easy solutions to systemic issues, 
activists might use AR 
to achieve justice-oriented outcomes. These outcomes can be characterized by goals of transformation, recognition, reciprocity, enablement, distribution, and accountability~\cite{dombrowski2016_4}. 
Practitioners' use of AR along the trajectories of these goals do this work through both digital media and non-digital social justice practices. The effort is heterotopic, as are the outcomes, which can exist in both physical and lived spaces as well as those of media representation, AR or otherwise \cite{schranz2014_12}.

For example, the Movers and Shakers \cite{Kinfolk_18} produced and developed The Monument App (2021) to place AR statues of important women, people of color, and LGBTQ+ icons in places of prominence around New York City (Fig. \ref{fig:kinfolk}). Additionally, they released a curriculum to be used with public schools in the city along with the mobile app. The work’s outcome provides reciprocity (recompense for those that deserve justice); and enablement (facilitating the growth of people to develop their own capacities ~\cite{hidalgo2015_7}) in its effort to decolonize the history of the city. 

Advocates using AR for social justice may also understand their outcomes through the lens of spatial justice, or “the fair and equitable distribution in the space of socially valued resources and the opportunities to utilize them"~\cite{soja2013_16}, particularly for supporting liberatory narratives and experiences. 
Outside of AR, digital technologies like games have been used for liberatory speculative design, mediating and generating new realities~\cite{coulton2016games,tynes2023toward,coopilton2022critical}. 
Marginalized communities could similarly use AR to alter the physical world with a new layer of reality and imagine liberatory places and futures, developing visions and strategies of how they can manifest and shape them. Communities could use AR to explore answers to questions like 
“what could this building become if we repurposed it to empower our community as technology creators,” or “how would we build and sustain a utopia where we are free from oppression?”

We invite workshop participants to speculate on the future outcomes of AR for spatial justice, and the actualization of digital efforts into tangible outcomes, physical or otherwise. As part of this process, scholars and practitioners should consider: accessibility, i.e., how to achieve outcomes when augmentations may not be cohesive or shared across community populations~\cite{Manfredo2023_10}; sustainability, as activists need to grapple with corporate infrastructures that support ubiquitous AR~\cite{Manfredini2019-hh}; and community engagement regarding who is involved in these AR experiences and how they engage with them~\cite{barba2011_1,oh2016_11}.

\section{Organizers}
The organizers have a diverse range of experience in creating and researching AR projects for social justice. Their combined expertise includes scholarship in different academic disciplines, work at industry leaders in AR, membership in and partnerships with historically marginalized communities, and editorship of a book on \textit{Augmented and Mixed Reality for Communities} \cite{fisher2021augmented}. They also have a track record of running successful workshops (e.g., Digital Wellbeing at CHI 2019 \cite{Cecchinato2019-td} and Dark Patterns at CHI 2021 \cite{Lukoff2021-ru}). Most of the organizers are early career researchers and all are eager to build community around this emerging area of research and practice.

\href{https://rafaelmlsilva.com}{Rafael M. L. Silva} (corresponding organizer) is a PhD candidate in Human Centered Design and Engineering at the University of Washington. His research explores the design, use, and social contextualization of emerging technologies (i.e., IoT and Immersive Media). He is particularly interested in understanding how technology for civic engagement can be leveraged by people in their everyday lives in order to build more just futures.

\href{https://amcard.myportfolio.com/}{Ana María Cárdenas Gasca} is a PhD student of the Expressive Computation Lab at the University of California Santa Barbara. Her research focuses on developing, informing, and studying technologies for spatial storytelling in the context of memorialization and documentation of Human Rights violations. She has collaborated with the Museo de la Memoria in Colombia, exploring AR applications for presenting victims' testimonials.

\href{https://www.jafisherportfolio.com/}{Joshua A. Fisher} is an assistant professor in the Center for Emerging Media Design and Development at Ball State University. His research focuses on utilizing emerging media for and with communities for interactive storytelling and expression. He is the XR Chair on the board for the Association for Research in Digital Interactive Narratives. 

\href{https://eripricru.wixsite.com/playfulresistance}{Erica Principe Cruz} is a PhD student of the Center for Transformational Play and Human-Computer Interaction Institute at Carnegie Mellon University. Her research explores how counterspaces that center the joy, rest, and healing of marginalized people can be co-created across realities. She is synthesizing strategies for cultivating counterspaces as games and AR/VR/XR experiences.

\href{https://www.linkedin.com/in/cinthyaejh/}{Cinthya Jauregui} is a masters student in Engineering Management and Leadership at Santa Clara University. Her work focuses on Human Computer Interaction. She is the Project Lead for the Thámien Ohlone AR Tour, where she facilitates co-designing with Muwekma Ohlone tribal leaders, humanities scholars, and computer science researchers to tell the story of Muwekma Ohlone past, present, and future through emerging AR technologies.

\href{https://amylueck.com/}{Amy J Lueck} is an associate professor of rhetoric and composition at Santa Clara University, where her research and teaching focus on histories of rhetorical instruction and practice, feminist historiography, cultural rhetorics, and rhetorical memory studies. Since 2018 she has been collaborating with Muwekma Ohlone and Ohlone tribal members on public-facing projects that use digital media to unsettle the patterns of Indigenous erasure that her research documents and to help sponsor the diverse cultural rhetorics practices of Ohlone youth.

\href{https://www.fannieliu.com}{Fannie Liu} is a VP Applied Research Lead on the Global Tech Applied Research AR/VR team at JPMorgan Chase \& Co. Her research involves the design of novel social experiences that leverage immersive technologies to promote communication and well-being. Previously, she was a Research Scientist at Snap, where she was the PI for research on the use of AR for activism. 

\href{https://www.andresmh.com/}{Andrés Monroy-Hernández} is an assistant professor in Princeton's Department of Computer Science and an associated faculty in Princeton's Center for Information Technology Policy. His research focuses on social computing, leveraging technologies such as AR and others. Previously, he led a research team at Snap focused on social AR.

\href{https://kailukoff.com/}{Kai Lukoff} is an assistant professor of Computer Science and Engineering at Santa Clara University. He is part of an interdisciplinary team of Muwekma Ohlone tribal leaders, humanities scholars, and computer science researchers who are creating an AR walking tour of the Native American history of Mission Santa Clara. He is developing a toolkit of AR resources that empower educators and storytellers around the world to develop ‘counter-tours’ that challenge hegemonic narratives of cultural heritage sites.

\section{Publication of Workshop Proceedings}

Accepted workshop papers will be published in \href{https://arxiv.org}{arXiv} arXiv, an open-access, online repository of research papers. This follows the recommendation to publish position papers in the field of computer science \cite{Aaron_6} and the practice of previous workshop proceedings published at the CHI Conference that have used report numbers \cite{fuentes2019_5,tigwell2019_17}. This will ensure a stable archive that makes it easy for  scholars to find and reference the workshop papers in the future.

\section{Workshop Activities}
The core activities of this workshop are interactive sessions in small- and medium-sized groups (Table \ref{tab:1}). The workshop will begin with a round of lightning introductions. In Session 1 (Artifact discussions), each participant will have a chance to share and discuss their artifact in a small group. In Session 2 (Working across boundaries), medium-sized groups will discuss: (a) Interdisciplinary collaboration;
(b) Partnering with community groups; (c) Navigating site permissions; and (d) Additional topics suggested by participants. In Session 3 (Accessibility, sustainability, and impact), medium-sized groups will discuss (a) Democratizing creator tools; (b) Scaling and maintaining impact; (c) Designing an artifact database; and (d) Additional topics suggested by participants. Instead of one-way presentations, workshop activities are intended to facilitate active dialogue between participants.

\section{Hybrid Format and Asynchronous Engagement}
This workshop will be held as a half-day hybrid workshop with 25-35 participants. This format was selected in order to enable meaningful participation from a wide audience, including those who are unable to travel to Hawaii or have chosen not to travel due to environmental or social justice concerns.  Organizers are committed to creating a rewarding experience with interaction between all participants.

There will be a shared online document with a section for each group in each workshop session. This section will include: group name, participants, links to relevant materials (e.g., artifacts), three initial discussion questions proposed by the group host, and space for note-taking. These collaborative  notes serve as: (1) a record of the workshop discussions; and (2) asynchronous materials in case technical or accessibility issues prohibit synchronous engagement.

Participants will be assigned to groups in advance of the workshop based on their preferences in a pre-workshop survey. This will enable the organizers to accommodate preferences, manage group size, and arrange the physical conference room accordingly. For small groups (about 4-participants), we will assign groups to different corners of the room and request they use one laptop per group to facilitate discussion between in-person and virtual participants. For medium groups (about 8-participants), we plan to set up four tables in the room, each with a single laptop connected to a wide angle webcam and an external monitor. Virtual participants will be shown on the external monitor. In-person participants will sit in a semi-circle so that they can be seen by the wide angle webcam and view the monitor.

A pre-workshop survey will ask participants about accessibility requirements (e.g., transcription). Organizers will plan the workshop around these requirements and request support from the CHI Workshop Chairs as necessary.

\begin{table*}[htbp]
    \centering
\rowcolors{2}{LightCyan}{white}
\begin{tabular}{p{0.05\textwidth}p{0.25\textwidth}p{0.6\textwidth}}
\rowcolor{gray!25}
Time & Activity & Description \\
\hline
9:00 & Welcome &
Organizers introduce themselves and the agenda \\ 
9:15 &
Lightning introductions &
Participants introduce themselves and share one slide with an example of an artifact from a social justice AR experience that connects to their work (one-minute per participant) \\ 
9:45 & Session 1A: Artifact discussions & 
Breakout rooms with four participants. One participant serves as a host and briefly presents their artifact (about 3-minutes) for discussion (about 12-minutes). After sessions 1A-1D, each participant will have served as the host for one breakout room. 
\\
10:00 & Short break & \\  
10:10 & Session 1B: Artifact discussions & \\ 
10:25 & Session 1C: Artifact discussions & \\
10:40 & Session 1D: Artifact discussions & \\ 
10:55 & Coffee break & \\ 
11:15 & Session 2: Working across boundaries &
Group discussions (about 8 participants) \\ 
11:45 & Group sharing & Rapporteur from each group shares Session 2 highlights \\ 
11:55 & Short break & \\ 
12:05 &
Session 3: Accessibility, sustainability, and impact &
Group discussions (about 8 participants) \\ 
12:35 & Group sharing &
Rapporteur from each group shares Session 3 highlights \\ 
12:50 & Next steps &
Conclusion and sign up participants to contribute to post-workshop plans \\ 
12:55 & End of formal activities & \\ 
13:00 & In-person social event &
Optional lunch \\ 
& Virtual social event &
Optional online hangout \\ 
\end{tabular}
    \caption{Draft schedule for the workshop}
    \label{tab:1}
\end{table*}

\section{Post-Workshop Plans}


The primary outcome of this workshop will be an interactive collection of site-specific social justice AR projects. Each entry will feature \textit{speculative artifact(s), process artifact(s)}, and/or \textit{final artifacts}. At present, a major challenge for designers is the lack of awareness of past work in site-specific AR. Such experiences are extremely difficult to access because (1) they require the user to be in a specific location; and (2) the technology behind AR is rapidly changing, so a platform or device that is supported today may not be supported tomorrow.

The starting point for this collection will be the annotated artifacts that workshop authors are required to include as a part of their submission. At the workshop, participants will be encouraged to identify further projects and artifacts to include, as well as relevant tags to explore this collection. After the workshop, the workshop organizers and interested participants will create a public website that highlights these projects. Users will be able to explore projects on a global map and via category tags such as technology (e.g., webAR, iOS app, Android app), site information (e.g, museum, monument, neighborhood), and identity (e.g., Black, Indigenous, women), and project metadata (e.g., Authors, Venue, Social Media). New entries will be accepted via an online form. This online collection is inspired by the Locomotion Vault \cite{di2021_3} and Haptipedia \cite{seifi2019_13,seifi2020_14}, community-sourced interactive databases that have resulted in CHI publications and serve as a resource for the wider community.

To disseminate the results of the workshop, the organizers will also publish a blog post of highlights on the UX Collective (\href{https://uxdesign.cc/}{uxdesign.cc}), a Medium publication with over 400,000 followers, mostly design practitioners. One of the organizers on this proposal did this for a previous workshop at CHI 2019 and it has served as a popular summary of the workshop discussion \cite{lukoff2019_9}. A similar blog post will share a summary of our discussions about AR for social justice.



\section{Call for Participation}

This half-day hybrid workshop will bring together an interdisciplinary group of researchers to discuss site-specific augmented reality (AR) for social justice. AR offers unique affordances, such as a strong tie to the physical environment and an ability to circumvent certain permissions when needed, that align with many forms of activism. Yet open questions remain about why, when, and how activists should leverage AR for social justice.

Applicants should submit a 1-6 page position paper (excluding references) as a single-column manuscript using the \href{https://chi2024.acm.org/submission-guides/chi-publication-formats/}{ACM Conference Proceedings Primary Article Template}. Submissions should start by presenting one specific artifact. This could be a \textit{speculative artifact}, such as a sketch or diagram of an imagined AR experience. It could be a \textit{process artifact}, such as a labeled site map or early prototype. Or it might be a \textit{final artifact}, such as a soundscape or a video clip of a completed experience. Authors may select and discuss an artifact created by someone else, it does not have to be their own work.

The remainder of the position paper should discuss the artifact as it relates to the themes of the workshop and prior work. The artifacts will not only guide our discussions at the workshop but also form the first entries in an anticipated interactive database of social justice AR resources that will be created as an outcome of the workshop.

Submissions will be reviewed by the workshop organizers based on quality, relevance, and diversity. Accepted papers will be published as workshop proceedings on arXiv. At least one author of each accepted submission must attend the workshop and all participants must register for both the workshop and at least one day of the conference.

\begin{itemize}
\item Website: \href{https://ar4socialjustice.org}{https://ar4socialjustice.org}
\item Date: Sunday, 12 May, 2024
\item Time: 9am-1pm HST (Hawaii Standard Time) (tbc)
\item Format: Hybrid (in-person and virtual)
\item Submit applications to: \href{https://forms.gle/e19BiRsG1tRCHQCAA}{http://tinyurl.com/ar4socialjustice}
\item Submission deadline: 1 March, 2024
\item Acceptance notification: 22 March, 2024
\end{itemize}

\begin{acks}
The opinions, findings, and conclusions or recommendations expressed in this material are those of the author(s) and do not necessarily reflect the views of JPMorgan Chase \& Co. or its affiliates. This paper was prepared for informational purposes with contributions from the Global Technology Applied Research center of JPMorgan Chase \& Co. This paper is not a product of the Research Department of JPMorgan Chase \& Co. or its affiliates. Neither JPMorgan Chase \& Co. nor any of its affiliates makes any explicit or implied representation or warranty and none of them accept any liability in connection with this paper, including, without limitation, with respect to the completeness, accuracy, or reliability of the information contained herein and the potential legal, compliance, tax, or accounting effects thereof. This research was funded in part by the NSF ER2-Ethical and Responsible Research Program (Award: 2026286).
\end{acks}

\newpage


\bibliographystyle{ACM-Reference-Format}
\bibliography{citation}


\begin{thebibliography}{35}


\ifx \showCODEN    \undefined \def \showCODEN     #1{\unskip}     \fi
\ifx \showDOI      \undefined \def \showDOI       #1{#1}\fi
\ifx \showISBNx    \undefined \def \showISBNx     #1{\unskip}     \fi
\ifx \showISBNxiii \undefined \def \showISBNxiii  #1{\unskip}     \fi
\ifx \showISSN     \undefined \def \showISSN      #1{\unskip}     \fi
\ifx \showLCCN     \undefined \def \showLCCN      #1{\unskip}     \fi
\ifx \shownote     \undefined \def \shownote      #1{#1}          \fi
\ifx \showarticletitle \undefined \def \showarticletitle #1{#1}   \fi
\ifx \showURL      \undefined \def \showURL       {\relax}        \fi
\providecommand\bibfield[2]{#2}
\providecommand\bibinfo[2]{#2}
\providecommand\natexlab[1]{#1}
\providecommand\showeprint[2][]{arXiv:#2}

\bibitem[Abebe et~al\mbox{.}(2022)]%
        {Abebe2022-nq}
\bibfield{author}{\bibinfo{person}{Veronica Abebe}, \bibinfo{person}{Gagik Amaryan}, \bibinfo{person}{Marina Beshai}, \bibinfo{person}{{Ilene}}, \bibinfo{person}{Ali~Ekin Gurgen}, \bibinfo{person}{Wendy Ho}, \bibinfo{person}{Naaji~R Hylton}, \bibinfo{person}{Daniel Kim}, \bibinfo{person}{Christy Lee}, \bibinfo{person}{Carina Lewandowski}, \bibinfo{person}{Katherine~T Miller}, \bibinfo{person}{Lindsey~A Moore}, \bibinfo{person}{Rachel Sylwester}, \bibinfo{person}{Ethan Thai}, \bibinfo{person}{Frelicia~N Tucker}, \bibinfo{person}{Toussaint Webb}, \bibinfo{person}{Dorothy Zhao}, \bibinfo{person}{Haicheng~Charles Zhao}, {and} \bibinfo{person}{Janet Vertesi}.} \bibinfo{year}{2022}\natexlab{}.
\newblock \showarticletitle{{Anti-Racist} {HCI}: notes on an emerging critical technical practice}. In \bibinfo{booktitle}{\emph{{CHI} Conference on Human Factors in Computing Systems Extended Abstracts}} (New Orleans, LA, USA) \emph{(\bibinfo{series}{CHI EA '22}, \bibinfo{number}{Article 1})}. \bibinfo{publisher}{Association for Computing Machinery}, \bibinfo{address}{New York, NY, USA}, \bibinfo{pages}{1--12}.
\newblock
\showISBNx{9781450391566}
\urldef\tempurl%
\url{https://doi.org/10.1145/3491101.3516382}
\showDOI{\tempurl}


\bibitem[Barba and MacIntyre(2011)]%
        {barba2011_1}
\bibfield{author}{\bibinfo{person}{Evan Barba} {and} \bibinfo{person}{Blair MacIntyre}.} \bibinfo{year}{2011}\natexlab{}.
\newblock \showarticletitle{A scale model of mixed reality}. In \bibinfo{booktitle}{\emph{Proceedings of the 8th ACM Conference on Creativity and Cognition}}. \bibinfo{publisher}{ACM Press}, \bibinfo{address}{New York}, \bibinfo{pages}{117--126}.
\newblock
\urldef\tempurl%
\url{https://doi.org/10.1145/2069618.2069640}
\showDOI{\tempurl}


\bibitem[Bolter et~al\mbox{.}(2021)]%
        {Bolter2021-bj}
\bibfield{author}{\bibinfo{person}{Jay~David Bolter}, \bibinfo{person}{Maria Engberg}, {and} \bibinfo{person}{Blair MacIntyre}.} \bibinfo{year}{2021}\natexlab{}.
\newblock \bibinfo{booktitle}{\emph{Reality Media: Augmented and Virtual Reality}}.
\newblock \bibinfo{publisher}{MIT Press}.
\newblock
\showISBNx{9780262361927}
\urldef\tempurl%
\url{https://play.google.com/store/books/details?id=VjYXEAAAQBAJ}
\showURL{%
\tempurl}


\bibitem[C{\'a}rdenas~Gasca et~al\mbox{.}(2022)]%
        {cardenas2022_2}
\bibfield{author}{\bibinfo{person}{Ana~Mar{\'\i}a C{\'a}rdenas~Gasca}, \bibinfo{person}{Jennifer~Mary Jacobs}, \bibinfo{person}{Andr{\'e}s Monroy-Hern{\'a}ndez}, {and} \bibinfo{person}{Michael Nebeling}.} \bibinfo{year}{2022}\natexlab{}.
\newblock \showarticletitle{AR Exhibitions for Sensitive Narratives: Designing an Immersive Exhibition for the Museum of Memory in Colombia}.
\newblock \bibinfo{journal}{\emph{Designing Interactive Systems Conference}} (\bibinfo{year}{2022}), \bibinfo{pages}{1698--1714}.
\newblock
\urldef\tempurl%
\url{https://doi.org/10.1145/3532106.3533549}
\showDOI{\tempurl}


\bibitem[Cecchinato et~al\mbox{.}(2019)]%
        {Cecchinato2019-td}
\bibfield{author}{\bibinfo{person}{Marta~E Cecchinato}, \bibinfo{person}{John Rooksby}, \bibinfo{person}{Alexis Hiniker}, \bibinfo{person}{Sean Munson}, \bibinfo{person}{Kai Lukoff}, \bibinfo{person}{Luigina Ciolfi}, \bibinfo{person}{Anja Thieme}, {and} \bibinfo{person}{Daniel Harrison}.} \bibinfo{year}{2019}\natexlab{}.
\newblock \showarticletitle{Designing for Digital Wellbeing: A Research \& Practice Agenda}. In \bibinfo{booktitle}{\emph{Extended Abstracts of the 2019 {CHI} Conference on Human Factors in Computing Systems}} (Glasgow, Scotland Uk) \emph{(\bibinfo{series}{CHI EA '19}, \bibinfo{number}{Paper W17})}. \bibinfo{publisher}{Association for Computing Machinery}, \bibinfo{address}{New York, NY, USA}, \bibinfo{pages}{1--8}.
\newblock
\showISBNx{9781450359719}
\urldef\tempurl%
\url{https://doi.org/10.1145/3290607.3298998}
\showDOI{\tempurl}


\bibitem[Coopilton(2022)]%
        {coopilton2022critical}
\bibfield{author}{\bibinfo{person}{Matthew Coopilton}.} \bibinfo{year}{2022}\natexlab{}.
\newblock \showarticletitle{Critical Game Literacies and Critical Speculative Imagination: A Theoretical and Conceptual Review}.
\newblock \bibinfo{journal}{\emph{gamevironments}} \bibinfo{number}{17} (\bibinfo{year}{2022}), \bibinfo{pages}{51--51}.
\newblock


\bibitem[Costanza-Chock(2020)]%
        {costanza2020design}
\bibfield{author}{\bibinfo{person}{Sasha Costanza-Chock}.} \bibinfo{year}{2020}\natexlab{}.
\newblock \bibinfo{booktitle}{\emph{Design justice: Community-led practices to build the worlds we need}}.
\newblock \bibinfo{publisher}{The MIT Press}.
\newblock


\bibitem[Coulton et~al\mbox{.}(2016)]%
        {coulton2016games}
\bibfield{author}{\bibinfo{person}{Paul Coulton}, \bibinfo{person}{Dan Burnett}, {and} \bibinfo{person}{Adrian Gradinar}.} \bibinfo{year}{2016}\natexlab{}.
\newblock \showarticletitle{Games as speculative design: Allowing players to consider alternate presents and plausible futures}. In \bibinfo{booktitle}{\emph{Proceedings of DRS 2016, Design Research Society 50th Anniversary Conference}}.
\newblock


\bibitem[Di~Luca et~al\mbox{.}(2021)]%
        {di2021_3}
\bibfield{author}{\bibinfo{person}{Massimiliano Di~Luca}, \bibinfo{person}{Hasti Seifi}, \bibinfo{person}{Simon Egan}, {and} \bibinfo{person}{Mar Gonzalez-Franco}.} \bibinfo{year}{2021}\natexlab{}.
\newblock \showarticletitle{Locomotion vault: the extra mile in analyzing vr locomotion techniques}. In \bibinfo{booktitle}{\emph{Proceedings of the 2021 CHI Conference on Human Factors in Computing Systems}}. \bibinfo{publisher}{ACM Press}, \bibinfo{address}{New York}, \bibinfo{pages}{1--10}.
\newblock
\urldef\tempurl%
\url{https://doi.org/10.1145/3411764.3445319}
\showDOI{\tempurl}


\bibitem[Dombrowski et~al\mbox{.}(2016)]%
        {dombrowski2016_4}
\bibfield{author}{\bibinfo{person}{Lynn Dombrowski}, \bibinfo{person}{Ellie Harmon}, {and} \bibinfo{person}{Sarah Fox}.} \bibinfo{year}{2016}\natexlab{}.
\newblock \showarticletitle{Social justice-oriented interaction design: Outlining key design strategies and commitments}. In \bibinfo{booktitle}{\emph{Proceedings of the 2016 ACM Conference on Designing Interactive Systems}}. \bibinfo{publisher}{ACM Press}, \bibinfo{address}{New York}, \bibinfo{pages}{656--671}.
\newblock
\urldef\tempurl%
\url{https://doi.org/10.1145/2901790.2901861}
\showDOI{\tempurl}


\bibitem[Fisher(2021)]%
        {fisher2021augmented}
\bibfield{author}{\bibinfo{person}{Joshua~A Fisher}.} \bibinfo{year}{2021}\natexlab{}.
\newblock \bibinfo{booktitle}{\emph{Augmented and Mixed Reality for Communities}}.
\newblock \bibinfo{publisher}{CRC Press}.
\newblock


\bibitem[Fuentes et~al\mbox{.}(2019)]%
        {fuentes2019_5}
\bibfield{author}{\bibinfo{person}{Carolina Fuentes}, \bibinfo{person}{Martin Porcheron}, \bibinfo{person}{Joel Fischer}, \bibinfo{person}{Nervo Verdezoto}, \bibinfo{person}{Oren Zuckerman}, \bibinfo{person}{Enrico Constanza}, \bibinfo{person}{Valeria Herskovic}, {and} \bibinfo{person}{Leila Takayama}.} \bibinfo{year}{2019}\natexlab{}.
\newblock \showarticletitle{Proceedings of the CHI 2019 Workshop on New Directions for the IoT: Automate, Share, Build, and Care}.
\newblock \bibinfo{journal}{\emph{arXiv preprint arXiv:1906.06089}} (\bibinfo{year}{2019}).
\newblock
\urldef\tempurl%
\url{http://arxiv.org/abs/1906.06089}
\showURL{%
\tempurl}


\bibitem[Fujihata({[n.\,d.]})]%
        {fujihataBeHere1942New}
\bibfield{author}{\bibinfo{person}{Masaki Fujihata}.} \bibinfo{year}{[n.\,d.]}\natexlab{}.
\newblock \bibinfo{title}{{{BeHere}}\hspace{0.166em}/\hspace{0.166em}1942:~{{A New Lens}} on the {{Japanese American Incarceration}}\textemdash{{BeHere App}} | {{Japanese American National Museum}}}.
\newblock \bibinfo{howpublished}{https://www.janm.org/exhibits/behere1942/app}.
\newblock


\bibitem[Hahn(2020)]%
        {BLAM}
\bibfield{author}{\bibinfo{person}{Jennifer Hahn}.} \bibinfo{year}{2020}\natexlab{}.
\newblock \bibinfo{title}{BLAM app lets users erect augmented reality statues of historical black figures}.
\newblock
\newblock
\urldef\tempurl%
\url{https://www.dezeen.com/2020/11/06/blam-history-bites-black-month-app-augmented-reality/}
\showURL{%
Retrieved October 10, 2023 from \tempurl}


\bibitem[Hertzmann(2023)]%
        {Aaron_6}
\bibfield{author}{\bibinfo{person}{Aaron Hertzmann}.} \bibinfo{year}{2023}\natexlab{}.
\newblock \showarticletitle{Computer Science Venues Should Publish Position Papers}.
\newblock \bibinfo{journal}{\emph{Aaron Hertzmann’s blog}} (\bibinfo{year}{2023}).
\newblock
\urldef\tempurl%
\url{https://aaronhertzmann.com/2023/06/30/meta-papers.html}
\showURL{%
Retrieved October 2, 2023 from \tempurl}


\bibitem[Hidalgo(2015)]%
        {hidalgo2015_7}
\bibfield{author}{\bibinfo{person}{LeighAnna Hidalgo}.} \bibinfo{year}{2015}\natexlab{}.
\newblock \showarticletitle{Augmented fotonovelas: Creating new media as pedagogical and social justice tools}.
\newblock \bibinfo{journal}{\emph{Qualitative Inquiry: QI}} \bibinfo{volume}{21}, \bibinfo{number}{3} (\bibinfo{year}{2015}), \bibinfo{pages}{300--314}.
\newblock
\urldef\tempurl%
\url{https://doi.org/10.1177/1077800414557831}
\showDOI{\tempurl}


\bibitem[{instute of museum {and} library services}({[n.\,d.]})]%
        {instuteofmuseumandlibraryservicesRememberingEmmettTill}
\bibfield{author}{\bibinfo{person}{{instute of museum {and} library services}}.} \bibinfo{year}{[n.\,d.]}\natexlab{}.
\newblock \bibinfo{title}{Remembering {{Emmett Till}}: {{From Chicago}} to {{Mississippi}}, {{Connecting Visitors}} with {{Location-Based History}}}.
\newblock \bibinfo{howpublished}{http://www.imls.gov/grant-spotlights/remembering-emmett-till-chicago-mississippi-connecting-visitors-location-based}.
\newblock


\bibitem[Kinfolk(2017)]%
        {Kinfolk_18}
\bibfield{author}{\bibinfo{person}{Kinfolk}.} \bibinfo{year}{2017}\natexlab{}.
\newblock \bibinfo{title}{Brewster, Idris; Cantave, Glenn}.
\newblock
\newblock
\urldef\tempurl%
\url{https://kinfolkhistory.com/}
\showURL{%
Retrieved October 7, 2023 from \tempurl}


\bibitem[Lukoff(2023)]%
        {Lukoff2023-do}
\bibfield{author}{\bibinfo{person}{Kai Lukoff}.} \bibinfo{year}{2023}\natexlab{}.
\newblock \bibinfo{title}{{Co-Designing} the Th{\'a}mien Ohlone Augmented Reality Tour at Santa Clara University}.
\newblock
\newblock
\urldef\tempurl%
\url{https://www.youtube.com/watch?v=ZawALZuiRIg}
\showURL{%
\tempurl}


\bibitem[Lukoff et~al\mbox{.}(2021)]%
        {Lukoff2021-ru}
\bibfield{author}{\bibinfo{person}{Kai Lukoff}, \bibinfo{person}{Alexis Hiniker}, \bibinfo{person}{Colin~M Gray}, \bibinfo{person}{Arunesh Mathur}, {and} \bibinfo{person}{Shruthi Chivukula}.} \bibinfo{year}{2021}\natexlab{}.
\newblock \showarticletitle{What Can {CHI} Do About Dark Patterns?}. In \bibinfo{booktitle}{\emph{Extended Abstracts of the 2021 {CHI} Conference on Human Factors in Computing Systems}} (Yokohama, Japan). \bibinfo{publisher}{ACM}, \bibinfo{address}{New York, NY, USA}.
\newblock
\urldef\tempurl%
\url{https://doi.org/10.1145/3411763.3441360}
\showDOI{\tempurl}


\bibitem[Lukoff and Munson(2019)]%
        {lukoff2019_9}
\bibfield{author}{\bibinfo{person}{Kai Lukoff} {and} \bibinfo{person}{Sean Munson}.} \bibinfo{year}{2019}\natexlab{}.
\newblock \showarticletitle{Digital wellbeing is way more than just reducing screen time}.
\newblock \bibinfo{journal}{\emph{UX Collective}} (\bibinfo{year}{2019}).
\newblock
\urldef\tempurl%
\url{https://uxdesign.cc/digital-wellbeing-more-than-just-reducing-screen-time-46223db9f057}
\showURL{%
Retrieved October 1, 2020 from \tempurl}


\bibitem[Manfredini(2019)]%
        {Manfredini2019-hh}
\bibfield{author}{\bibinfo{person}{Manfredo Manfredini}.} \bibinfo{year}{2019}\natexlab{}.
\newblock \showarticletitle{Simulation, Control and Desire: Urban Commons and {Semi-Public} Space Resilience in the Age of Augmented Transductive Territorial Production}.
\newblock \bibinfo{journal}{\emph{The journal of physiological sciences: JPS}} \bibinfo{volume}{4}, \bibinfo{number}{2} (\bibinfo{date}{Sept.} \bibinfo{year}{2019}), \bibinfo{pages}{179--198}.
\newblock
\showISSN{1880-6546, 2206-9658}
\urldef\tempurl%
\url{https://doi.org/10.32891/jps.v4i2.1209}
\showDOI{\tempurl}


\bibitem[Manfredini(2023)]%
        {Manfredo2023_10}
\bibfield{author}{\bibinfo{person}{Manfredo Manfredini}.} \bibinfo{year}{2023}\natexlab{}.
\newblock \showarticletitle{DISCUSSION AND CONCLUSION–SEMI-PUBLIC SPACE AND THE EMERGING SPATIALISATIONS OF RESILIENT URBAN COMMONS}.
\newblock \bibinfo{journal}{\emph{Discussion and Conclusion - Semi-Public Space and the Emerging Spatialisations of Resilient Urban Commons}} (\bibinfo{year}{2023}).
\newblock
\urldef\tempurl%
\url{https://www.drh.nz/files/2019/08/1.4.2-Discussion-and-conclusion-semi-public-space-commons.-Translocalism-Resilience.pdf}
\showURL{%
Retrieved October 7, 2023 from \tempurl}


\bibitem[McWhirter et~al\mbox{.}(2021)]%
        {McWhirter2021-yj}
\bibfield{author}{\bibinfo{person}{Joshua McWhirter}, \bibinfo{person}{Idris Brewster}, {and} \bibinfo{person}{Glenn Cantave}.} \bibinfo{year}{2021}\natexlab{}.
\newblock \bibinfo{title}{A Monumental Shift}.
\newblock \bibinfo{howpublished}{\url{https://www.guernicamag.com/a-monumental-shift/}}.
\newblock
\urldef\tempurl%
\url{https://www.guernicamag.com/a-monumental-shift/}
\showURL{%
\tempurl}
\newblock
\shownote{Accessed: 2023-10-10}.


\bibitem[Oh et~al\mbox{.}(2016)]%
        {oh2016_11}
\bibfield{author}{\bibinfo{person}{Soo~Youn Oh}, \bibinfo{person}{Ketaki Shriram}, \bibinfo{person}{Bireswar Laha}, \bibinfo{person}{Shawnee Baughman}, \bibinfo{person}{Elise Ogle}, {and} \bibinfo{person}{Jeremy Bailenson}.} \bibinfo{year}{2016}\natexlab{}.
\newblock \showarticletitle{Immersion at scale: Researcher's guide to ecologically valid mobile experiments}. In \bibinfo{booktitle}{\emph{2016 IEEE Virtual Reality (VR)}}. \bibinfo{publisher}{IEEE}, \bibinfo{address}{Greenville, SC, USA}, \bibinfo{pages}{249--250}.
\newblock
\urldef\tempurl%
\url{https://ieeexplore.ieee.org/abstract/document/7504747/}
\showURL{%
\tempurl}


\bibitem[Rucker(2021)]%
        {MonumentLab}
\bibfield{author}{\bibinfo{person}{Ursula Rucker}.} \bibinfo{year}{2021}\natexlab{}.
\newblock \bibinfo{title}{Self-guided tours of a public space by unearthing the multiple layers of history, meaning, and interpretation of that site through a personal smart device.}
\newblock
\newblock
\urldef\tempurl%
\url{https://monumentlab.com/projects/overtime}
\showURL{%
Retrieved October 10, 2023 from \tempurl}


\bibitem[Schranz(2014)]%
        {schranz2014_12}
\bibfield{author}{\bibinfo{person}{Christine Schranz}.} \bibinfo{year}{2014}\natexlab{}.
\newblock \showarticletitle{Augmented Reality in Design: Thinking about Hybrid Forms of Virtual and Physical Space in Design}. In \bibinfo{booktitle}{\emph{Design, User Experience, and Usability. User Experience Design for Diverse Interaction Platforms and Environments}}. \bibinfo{publisher}{Springer}, \bibinfo{address}{Switzerland}, \bibinfo{pages}{624--635}.
\newblock
\urldef\tempurl%
\url{https://doi.org/10.1007/978-3-319-07626-3_59}
\showDOI{\tempurl}


\bibitem[Schroeder et~al\mbox{.}(2023)]%
        {Schroeder2023-qn}
\bibfield{author}{\bibinfo{person}{Hope Schroeder}, \bibinfo{person}{Rob Tokanel}, \bibinfo{person}{Kyle Qian}, {and} \bibinfo{person}{Khoi Le}.} \bibinfo{year}{2023}\natexlab{}.
\newblock \showarticletitle{Location-based {AR} for Social Justice: Case Studies, Lessons, and Open Challenges}. In \bibinfo{booktitle}{\emph{Extended Abstracts of the 2023 {CHI} Conference on Human Factors in Computing Systems}} (Hamburg, Germany) \emph{(\bibinfo{series}{CHI EA '23}, \bibinfo{number}{Article 391})}. \bibinfo{publisher}{Association for Computing Machinery}, \bibinfo{address}{New York, NY, USA}, \bibinfo{pages}{1--10}.
\newblock
\showISBNx{9781450394222}
\urldef\tempurl%
\url{https://doi.org/10.1145/3544549.3573855}
\showDOI{\tempurl}


\bibitem[Seifi et~al\mbox{.}(2019)]%
        {seifi2019_13}
\bibfield{author}{\bibinfo{person}{Hasti Seifi}, \bibinfo{person}{Farimah Fazlollahi}, \bibinfo{person}{Michael Oppermann}, \bibinfo{person}{John~Andrew Sastrillo}, \bibinfo{person}{Jessica Ip}, \bibinfo{person}{Ashutosh Agrawal}, \bibinfo{person}{Gunhyuk Park}, \bibinfo{person}{Katherine~J Kuchenbecker}, {and} \bibinfo{person}{Karon~E MacLean}.} \bibinfo{year}{2019}\natexlab{}.
\newblock \showarticletitle{Haptipedia: Accelerating haptic device discovery to support interaction \& engineering design}. In \bibinfo{booktitle}{\emph{Proceedings of the 2019 CHI conference on human factors in computing systems}}. \bibinfo{publisher}{ACM Press}, \bibinfo{address}{New York}, \bibinfo{pages}{1--12}.
\newblock
\urldef\tempurl%
\url{https://doi.org/10.1145/3290605.3300788}
\showDOI{\tempurl}


\bibitem[Seifi et~al\mbox{.}(2020)]%
        {seifi2020_14}
\bibfield{author}{\bibinfo{person}{Hasti Seifi}, \bibinfo{person}{Michael Oppermann}, \bibinfo{person}{Julia Bullard}, \bibinfo{person}{Karon~E MacLean}, {and} \bibinfo{person}{Katherine~J Kuchenbecker}.} \bibinfo{year}{2020}\natexlab{}.
\newblock \showarticletitle{Capturing experts' mental models to organize a collection of haptic devices: Affordances outweigh attributes}. In \bibinfo{booktitle}{\emph{Proceedings of the 2020 CHI Conference on Human Factors in Computing Systems}}. \bibinfo{publisher}{ACM Press}, \bibinfo{address}{New York}, \bibinfo{pages}{1--12}.
\newblock
\urldef\tempurl%
\url{https://doi.org/10.1145/3313831.3376395}
\showDOI{\tempurl}


\bibitem[Silva et~al\mbox{.}(2022)]%
        {UnderstandingARactivism}
\bibfield{author}{\bibinfo{person}{Rafael M.~L. Silva}, \bibinfo{person}{Erica Principe~Cruz}, \bibinfo{person}{Daniela~K. Rosner}, \bibinfo{person}{Dayton Kelly}, \bibinfo{person}{Andr\'{e}s Monroy-Hern\'{a}ndez}, {and} \bibinfo{person}{Fannie Liu}.} \bibinfo{year}{2022}\natexlab{}.
\newblock \showarticletitle{Understanding AR Activism: An Interview Study with Creators of Augmented Reality Experiences for Social Change}. In \bibinfo{booktitle}{\emph{Proceedings of the 2022 CHI Conference on Human Factors in Computing Systems}} (New Orleans, LA, USA) \emph{(\bibinfo{series}{CHI '22})}. \bibinfo{publisher}{Association for Computing Machinery}, \bibinfo{address}{New York, NY, USA}, Article \bibinfo{articleno}{630}, \bibinfo{numpages}{15}~pages.
\newblock
\showISBNx{9781450391573}
\urldef\tempurl%
\url{https://doi.org/10.1145/3491102.3517605}
\showDOI{\tempurl}


\bibitem[Soja(2013)]%
        {soja2013_16}
\bibfield{author}{\bibinfo{person}{Edward~W Soja}.} \bibinfo{year}{2013}\natexlab{}.
\newblock \bibinfo{booktitle}{\emph{Seeking spatial justice}}. Vol.~\bibinfo{volume}{16}.
\newblock \bibinfo{publisher}{U of Minnesota Press}, \bibinfo{address}{Chicago}.
\newblock
\urldef\tempurl%
\url{https://play.google.com/store/books/details?id=NkfEeomy-IUC}
\showURL{%
\tempurl}


\bibitem[Tigwell et~al\mbox{.}(2019)]%
        {tigwell2019_17}
\bibfield{author}{\bibinfo{person}{Garreth~W Tigwell}, \bibinfo{person}{Zhanna Sarsenbayeva}, \bibinfo{person}{Benjamin~M Gorman}, \bibinfo{person}{David~R Flatla}, \bibinfo{person}{Jorge Goncalves}, \bibinfo{person}{Yeliz Yesilada}, {and} \bibinfo{person}{Jacob~O Wobbrock}.} \bibinfo{year}{2019}\natexlab{}.
\newblock \showarticletitle{Proceedings of the CHI'19 Workshop: Addressing the Challenges of Situationally-Induced Impairments and Disabilities in Mobile Interaction}.
\newblock \bibinfo{journal}{\emph{arXiv preprint arXiv:1904.05382}} (\bibinfo{year}{2019}).
\newblock
\urldef\tempurl%
\url{http://arxiv.org/abs/1904.05382}
\showURL{%
\tempurl}


\bibitem[Times(2020)]%
        {SkytypingAR}
\bibfield{author}{\bibinfo{person}{New~York Times}.} \bibinfo{year}{2020}\natexlab{}.
\newblock \bibinfo{title}{Protesting U.S. Immigration Policies, Artists Aim for the Sky}.
\newblock
\newblock
\urldef\tempurl%
\url{https://www.nytimes.com/2020/07/03/arts/design/july-4-skytyping-skywriting-immigration.html}
\showURL{%
Retrieved October 10, 2023 from \tempurl}


\bibitem[Tynes et~al\mbox{.}(2023)]%
        {tynes2023toward}
\bibfield{author}{\bibinfo{person}{Brendesha~M Tynes}, \bibinfo{person}{Matthew Coopilton}, \bibinfo{person}{Joshua Schuschke}, {and} \bibinfo{person}{Ashley Stewart}.} \bibinfo{year}{2023}\natexlab{}.
\newblock \showarticletitle{Toward Developmental Science That Meets the Challenges of 2044: Afrofuturist Development Theory, Design, and Praxis}.
\newblock In \bibinfo{booktitle}{\emph{Diversity and Developmental Science: Bridging the Gaps Between Research, Practice, and Policy}}. \bibinfo{publisher}{Springer}, \bibinfo{pages}{245--270}.
\newblock


\end{thebibliography}

\end{document}